\documentclass[12pt]{iopart}

\usepackage{iopams}  
\usepackage{graphicx}  
\usepackage{amsmath}
\newcommand{\ie}{{\it i.e.}\ }

\newcommand{\eg}{{\it e.g.}\ }

\newcommand{\ave}[1]{\left\langle#1 \right\rangle}

\newcommand{\elabel}[1]{\label{eq:#1}}
\newcommand{\seclabel}[1]{\label{sec:#1}}
\newcommand{\secref}[1]{Sec.~\ref{sec:#1}}

\newcommand{\flabel}[1]{\label{fig:#1}}

\newcommand{\be}{\begin{equation}}
\newcommand{\ee}{\end{equation}}
\newcommand{\bea}{\begin{eqnarray}}
\newcommand{\eea}{\end{eqnarray}}
\newcommand{\bc}{\begin{center}}
\newcommand{\ec}{\end{center}}

\begin{document}

\title[A recipe for irreproducible results]{A recipe for irreproducible results}

\author{Ole Peters$^{1,2}$, Maximilian J. Werner$^{1,3}$}
\address{$^1$London Mathematical Laboratory, 14 Buckingham Street, London WC2N 6DF, UK\\
$^2$Santa Fe Institute, 1399 Hyde Park Road, Santa Fe, NM 87501, USA\\
$^3$School of Earth Sciences and Cabot Institute, University of Bristol, Wills Memorial Building, Queen's Road, Bristol, BS8 1RJ, UK}
\ead{o.peters@lml.org.uk, max.werner@bristol.ac.uk}


\begin{abstract}
Recent studies have shown that many results published in peer-reviewed
scientific journals are not reproducible. This raises the following question: why is it so easy to fool myself into believing that a result is reliable when in fact it is not? Using Brownian motion as a toy model, we show how this can happen if ergodicity is assumed where it is unwarranted. A measured value can appear stable when judged over time, although it is not stable across the ensemble: a different result will be obtained each time the experiment is run.
\end{abstract}

%
%
%
%
%

\section{Introduction: The reproducibility crisis}
Science is in a reproducibility crisis -- a statement that 90\% of respondents to a survey by the journal Nature agreed with \cite{Baker2016}. More than 70\% of respondents stated that they had tried and failed to reproduce another scientist's experiments. A wide range of factors have been identified to contribute to irreproducible results, from insufficiently described protocols via misused statistics to biased reporting and outright fraud \cite{NatureEditorial2016}. 

In this paper we propose that irreproducibility of an entirely different (non-human) origin can arise in measurements of a particular class of systems. We construct a generic example of a statistical study that yields apparently solid, but in reality irreproducible results. The key idea is this: statistical studies often assume (explicitly or implicitly) that a quantity has a stationary true value. Measurements of the quantity are assumed to be distributed around that true value, according to a stationary distribution. To remove uncertainty in the measurement, the standard technique of time averaging is used to extract the true value. However, here we discuss how, if the stationarity assumption is false, time averaging can fool us in a subtle way: it will seem to work and yield an apparently reliable, temporally stable, value. But that value will not be reproducible: if we do the same experiment again, we will find a different result.

\subsection{Strategy}
To illustrate the pitfall we want to highlight, we will use the following strategy. 

1. Design a protocol to extract the true value from a noisy measurement. The protocol is designed to be good whenever a true value exists (and is stationary).

2. Apply the protocol inappropriately to the simplest case we can think of where no true value exists.

3. Observe that the protocol seems to work, although it actually just generates a random number (instead of the true value). 

Results are not reproducible: different runs of the experiment will give different results that -- crucially -- appear meaningful according to the protocol.

\subsection{Outline}
In \secref{Reproducibility} we give a high-level outline of the connection between reproducibility and ergodicity, as studied in dynamical systems and stochastic processes. This may be skipped by readers familiar with the issues.
In \secref{Measurement} we define the measurement protocol (step 1 in the strategy). 
In \secref{Model} we compute what will happen under this protocol if it is applied to an observable that follows Brownian motion (steps 2 and 3 in the strategy). This is illustrated numerically in \secref{Numerical}. Finally, in \secref{Discussion} we recap the results and argue that they boil down to a universal scaling argument. We list questions for further research.
\seclabel{}

\section{Reproducibility and ergodicity}
\seclabel{Reproducibility}
In this section we motivate concepts in ergodic and non-ergodic systems that are related to reproducibility. For readers interested in further details, reference \cite{Berger2001} discusses ergodicity from a dynamical systems perspective. The point of view in \cite{GrimmettStirzaker2001} is closer to stochastic processes. Key insights are often succinctly communicated in texts on numerical work, such as \cite{KloedenPlaten1999}. 

In dynamical systems we think of a phase space $\Omega$, each of whose elements, $\omega$, is a state of the system. A dynamical system traces out a trajectory, meaning it starts at initial state $\omega(0)$, then moves to $\omega(dt)$, then to $\omega(2dt)$ and so on. Dynamical systems are often thought of as deterministic, but there is a close connection to stochastic processes.
A good way to think of this connection is the example of a (pseudo-) random number generator. The initial state $\omega(0)$ is the state of the computer determined by the seed passed to the random number generator. From there on all future states of the computer are deterministic, though for all practical purposes they may be treated as stochastic.

\subsection{Ergodic dynamical systems}

In an ergodic system, irrespective of the initial state $\omega(t=0)$, any trajectory will eventually get arbitrarily close to any state in $\Omega$.

An observable, let's call it $a$, models something that can be measured. It is a function that associates a number to each state: if the system is in state $\omega$, then $a$ takes the value $a(\omega)$. Through the dynamic of the state $\omega(t)$, the observable depends on time, $a=a(\omega(t))$, see Fig.~\ref{fig:obs}. In the random-number-generator example, the observable would be the random number. The generator is some algorithm that marches the state of the computer, $\omega(t)$, forward, the observable output is a sequence of instances of a random number, $a(\omega(t))$, with a distribution requested by the user. Each instance of the random number is just the state of the computer mapped to the real line, $a=a(\omega(t))$.
\begin{figure}[h!]
\begin{picture}(200,300)(0,0)
  \put(80,-10){\includegraphics[width=.7\textwidth]{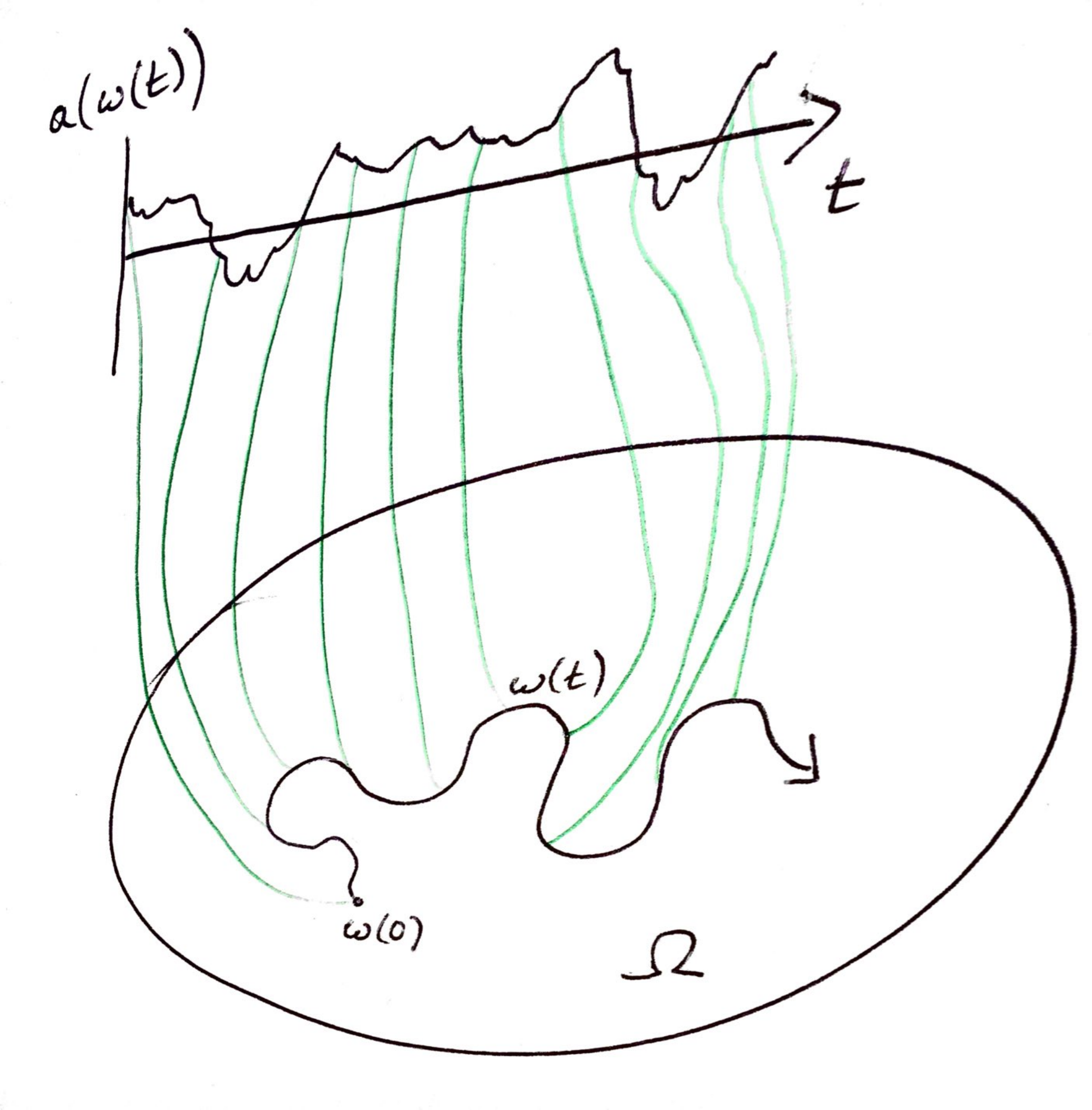}}
\end{picture}
\caption{A dynamical system generating a time series of observable $a(\omega(t))$.}
\flabel{obs}
\end{figure}

A measurement of an observable is imagined to be a slow process compared to the dynamic $\omega(t)$, so that the value generated by the measurement is really a (finite-)time average. For example, the pressure of a gas is a finite-time average over lots of fast particle collisions. Mathematically, 
\be
\overline{a}_t=\frac{1}{t}\int_0^t a(\omega(s)) ds.
\elabel{t_ave}
\ee
The averaging time $t$ is so large (the dynamic of $\omega$ is so fast) that each state is visited with a representative relative frequency. These frequencies constitute the invariant, or ergodic, probability measure, or simply the probability density, $P(\omega)$. The long-time limit of the finite-time average is denoted by dropping the subscript $t$, as $\lim_{t\to\infty}\overline{a}_t=\overline{a}$.
It coincides with the expectation value under the invariant measure, $\ave{a}_P=\int_\Omega P(\omega) a(\omega) d\omega$. We can also arrive at the expectation value by averaging an ensemble of $N$ values $\omega_i$ sampled from $P(\omega)$ 
\be
\ave{a}_{P, N}=\frac{1}{N}\sum_i^N a(\omega_i)
\ee
and taking the large-ensemble limit, $\lim_{N\to\infty}\ave{a}_{P, N}=\ave{a}_P$.
Ergodicity implies
\be
\overline{a}= \ave{a}_P.
\ee

Repeating an experiment is imagined as follows. \\
Experimental run 1: At the beginning of the first experiment, $t_1=0$, the system is in an unknown state $\omega_1(t_1=0)$. We measure $a$ as described by averaging over time.\\ 
Experimental run 2: At the beginning of the second experiment, $t_2=0$, and we start from a different unknown state $\omega_2(t_2=0)$. 

Provided that we measure for long enough, the ergodic property guarantees that we will arrive at the same measured value, namely $\ave{a}_P$, whenever we run the experiment, irrespective of the initial state. Uncertainties in the measured value vanish with the averaging time.

In this example, ergodicity thus guarantees reproducibility, provided the averaging time is sufficient.

\subsection{Strong ergodicity breaking}

Strong ergodicity breaking is a trivial way of destroying reproducibility. The phase space of a strongly non-ergodic system is separated into mutually inaccessible basins. In this case, two measurements of the same observable will not yield the same value if the (unknown) initial states for the two measurements, $\omega_1(0)$ and $\omega_2(0)$, are in different basins. Since $\omega(t=0)$ is considered unobservable and uncontrollable, experimental results will not be reproducible.
This is illustrated in Figure~\ref{fig:strong}.
\begin{figure}[h!]
\begin{picture}(200,230)(0,0)
  \put(60,-170){\includegraphics[width=.9\textwidth]{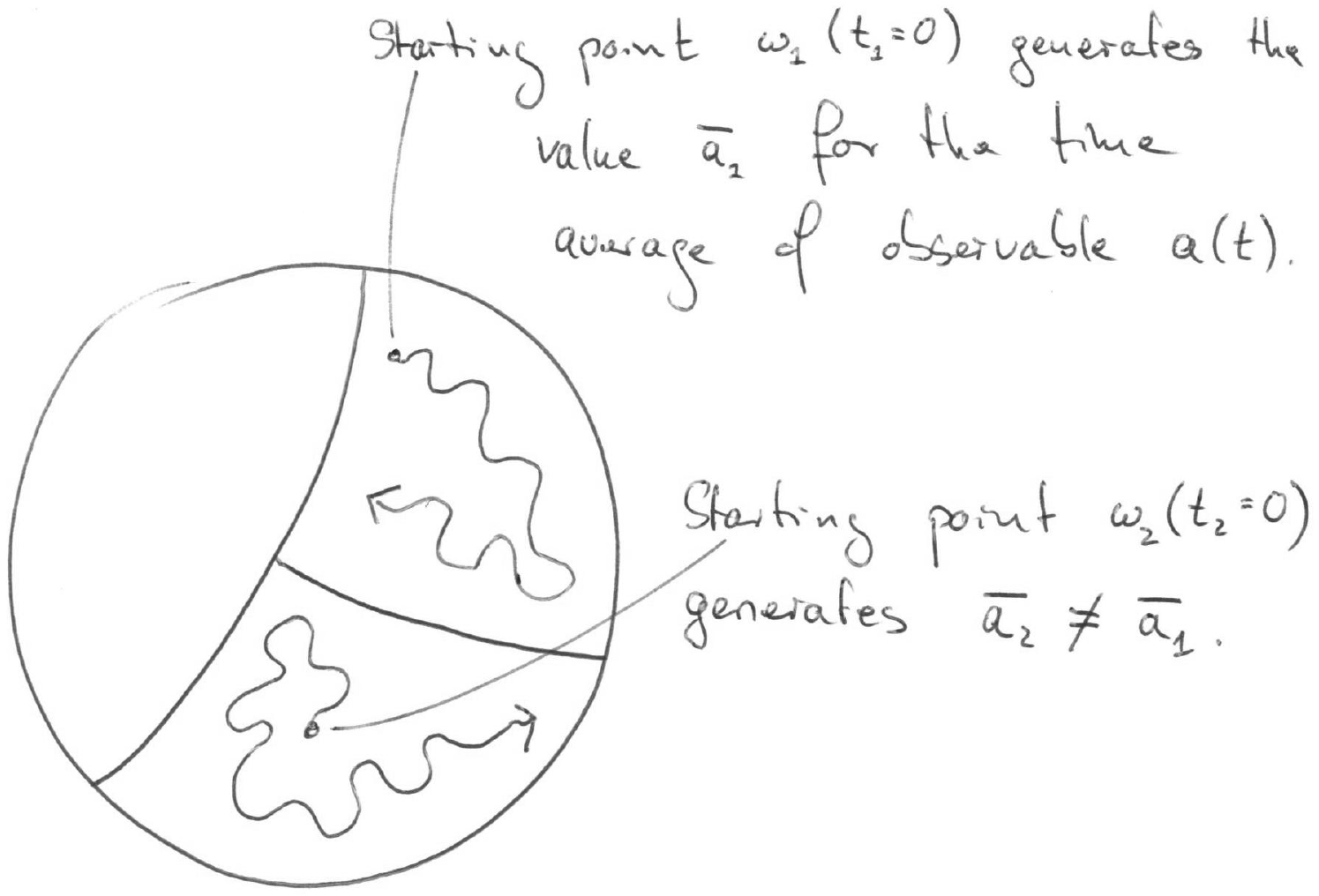}}
\end{picture}
\caption{In a strongly non-ergodic system, phase space is separated into mutually inaccessible basins. The time average of an observable will depend on the basin in which a realization starts. Thus, running the same experiment with different initial values will produce different results.}
\flabel{strong}
\end{figure}

\subsection{Weak ergodicity breaking and Brownian motion}
We will focus on a more subtle way of breaking ergodicity, called ``weak ergodicity breaking,'' that is typically studied in stochastic processes. This is often demonstrated with Polya's urn models, see \eg \cite[Chap. V (2)]{Feller1950}. While these urn models are good examples, they can give the wrong impression that the effect is unusual, or that one has to design a model specifically to display it. 
Instead of urn models, we will therefore work with a strongly universal model, namely Brownian motion  
\be
W(t)=\int_0^t dW,
\ee
where $dW$ is the Wiener increment, distributed normally with mean $0$ and variance $dt$, which we denote as $dW\sim \mathcal{N}(0,dt)$. The expectation value of Brownian motion is different from the time average because the time average does not converge to a number.

Brownian motion, being the sum of many Gaussian increments, is also distributed normally, as $W(t)\sim \mathcal{N}(0,t)$. It is important to bear in mind that the distribution is an ensemble property and does not immediately tell us much about temporal behavior. Specifically, the following distinction between ensemble behavior and temporal behavior must be made: the average over $N$ realizations of Brownian motion at some fixed time converges to zero as $N$ increases, 
\be
\frac{1}{N}\sum_i^N W_i(t) \sim \mathcal{N}\left(0,\frac{t}{N}\right).
\elabel{ens}
\ee 
However, the average up to time $t$ of a single Brownian motion does not converge to zero as $t$ increases. The expectation value (an ensemble property) of this finite-time average is zero, but the variance diverges with time, 
\be
\frac{1}{t}\int_0^t W(s) ds \sim \mathcal{N}\left(0,\frac{t}{3}\right).
\elabel{time}
\ee
A single Brownian trajectory and its time average are shown in Fig.~\ref{fig:trajectory}.

\begin{figure}[h!]
\begin{picture}(200,230)(0,0)
  \put(60,0){\includegraphics[width=.7\textwidth]{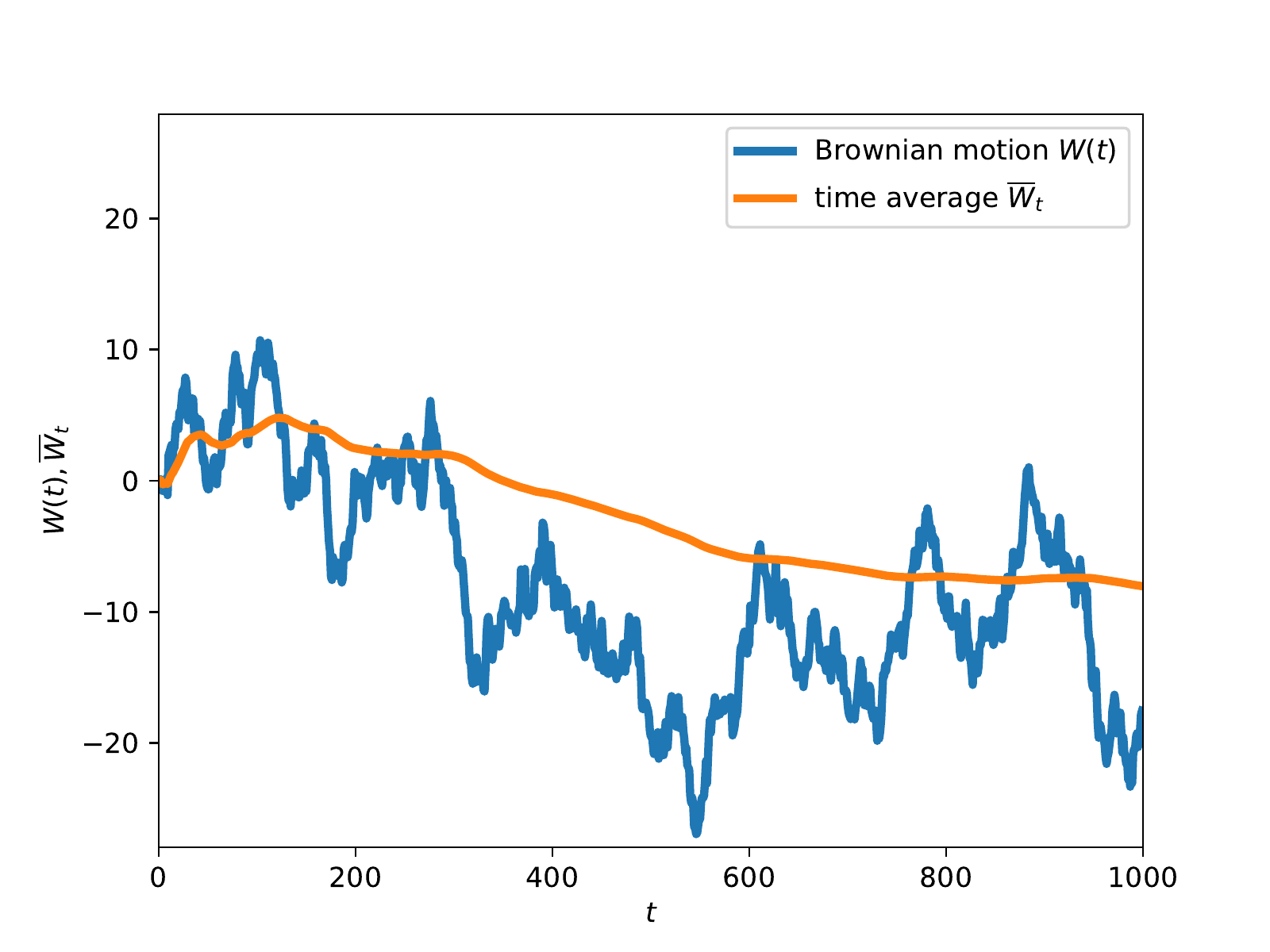}}
\end{picture}
\caption{Example trajectory of Brownian motion, $W(t)$, and its finite-time average, $\overline{W}_t$. Time averaging is a smoothing operation ensuring that $d\overline{W}_t/dt$ exists, although $dW/dt$ does not exist. Notice how variations in the time average diminish over time. The process $\overline{W}_t$ is sometimes called the random acceleration process.}
\flabel{trajectory}
\end{figure}

The limiting cases of Eq.~\ref{eq:ens} and Eq.~\ref{eq:time} couldn't be more different. The distribution in Eq.~\ref{eq:ens} in the limit $N\to\infty$ becomes infinitely peaked, \ie it converges to a Dirac delta function centered at zero. The distribution in Eq.~\ref{eq:time} in the limit $t\to\infty$ becomes infinitely broad, namely it will be zero everywhere while maintaining normalization.

These non-ergodic properties imply that the behavior of an ensemble of Brownian motion realizations is quite uninformative of the behavior of a single trajectory over time. Intuition developed in the ensemble setup is not valid in the single-trajectory setup.

\section{Measurement protocol}
\seclabel{Measurement}

In order to extract the ``true'' value from a sequence of measurements of the observable $a$, we compute the finite-time average, $\overline{a}_t$, according to eq.~\ref{eq:t_ave}.

We now design a criterion that will tell us when the time average has stabilized. ``Stabilized'' means that the estimate $\overline{a}_t$ doesn't change much any more. One way of determining that is to compute the square (or absolute value) of the rate of change $d\overline{a}_t /dt$ and insist that it be small over some time interval $\Delta t$. Thus, we're interested in
\bea
v(t;\Delta t)=\frac{1}{\Delta t} \int_{t-\Delta t}^{t} \left(  \left.\frac{d\overline{a}_\tau}{d\tau}\right|_{\tau=s} \right)^2 ds,
\elabel{variation}
\eea
which we will call the time variation.
Our criterion for stability will require this quantity to be smaller than some threshold $\epsilon$, so that the termination time $t^*$ is the first time when 
\be
v(t^*;\Delta t)<\epsilon.
\elabel{t_criterion}
\ee 
When the criterion is satisfied, the researcher concludes that the true value of $a$ is $\overline{a}_{t^*}\pm \Delta t \sqrt{v(t^*;\Delta t)}$. Note that this is an intuitive and  perfectly sensible way of proceeding if $a$ is observed in an ergodic system. We could invent many other stability criteria, but the simplest ones will resemble Eq.~\ref{eq:t_criterion} to some extent. Since $\Delta t$ is fixed in Eq.~\ref{eq:variation}, it is the rate of change of the time average, $\frac{d}{dt} \overline{a}_t$, that sets the termination time $t^*$.

\section{Model and key calculation}
\seclabel{Model}

Our protocol is indeed reasonable if $a$ is ergodic. But what if that's not the case? We will investigate the example of Brownian motion, $a(t)=W(t)$. The Brownian motion model is chosen because of its universality -- many real-world processes are well modelled by Brownian motion within some range of scales.  We will find below that the finite-time average of a Brownian motion, although it doesn't converge to anything, becomes smoother as time goes by and tricks the protocol: because of the increasing smoothness, the termination criterion will eventually be satisfied, and the researcher is fooled into believing that he has found, within some uncertainty band, the true value of the observable (which doesn't exist). Despite the apparently temporally stable single observation, repeating the experiment will yield a different result.

We want to know if the termination criterion will be satisfied: does the time average of a Brownian motion become indefinitely smoother over time? To answer this question we compute the rate of change of the time average, $\frac{d\overline{W}_t}{dt}$, square it, $\left(\frac{d\overline{W}_t}{dt}\right)^2$, and then take the expectation value, $\ave{\left(\frac{d\overline{W}_t}{dt}\right)^2}$. The expectation value is an average over (infinitely) many runs of the experiment. We postulate that it reflects the {\it typical} behavior of  $\frac{d\overline{W}_t}{dt}$. We believe this because of the square: in the expectation value of the square, outliers, \ie large positive or negative values, are weighted more heavily than in the expectation value of $\frac{d\overline{W}_t}{dt}$ itself. Therefore if the expectation value of the square converges to zero, there cannot be many a-typical trajectories that do not converge to zero. 

Below, we present exact results for the decay with time of the ensemble average of the {\it square} of the rate of change of the time-average. We confirmed via simulations that the ensemble average of the {\it absolute value} of the rate of change also decays to zero, as expected. 

$\overline{W}_t$ is the finite-time average of a Brownian motion, 
\be
\overline{W}_t=\frac{1}{t}\int_0^t W(s) ds.
\ee

We are interested in the time-derivative of $\overline{W}_t$,
\be
\frac{d}{dt} \overline{W}_t=-\frac{1}{t^2} \int_0^t W(s) ds + \frac{1}{t} W(t).
\ee
Note that this is the time derivative of the time average of a Brownian motion, not that of a Brownian 
motion itself (which does not exist).
We find the mean-square of this time derivative
\be
\ave{\left(\frac{d\overline{W}_t}{dt}\right)^2}=\ave{ \frac{1}{t^4} \left(\int_0^t W(s) ds \right)^2 +\frac{1}{t^2} W(t)^2  -\frac{2}{t^3}  W(t) \int_0^t W(s) ds}.
\elabel{mean-square}
\ee
Interestingly, the first two terms are trivial. The first term on the right-hand-side is the second moment of an integral over a Brownian motion, whose distribution is known, 
\be
\int_0^t W(s) ds \sim \mathcal{N}\left(0,\frac{1}{3} t^3\right).
\elabel{sum_dist}
\ee
The second term is the second moment of a Brownian motion itself, whose distribution is also known, 
\be
W(t)\sim \mathcal{N}(0,t). 
\ee 
Thus, Eq.~\ref{eq:mean-square} reduces to
\be
\ave{\left(\frac{d\overline{W}_t}{dt}\right)^2}= \frac{1}{3t} +\frac{1}{t}  -\frac{2}{t^3}  \ave{W(t) \int_0^t W(s) ds},
\elabel{mean-square-2}
\ee
and it only remains for us to evaluate the final term $\ave{W(t) \int_0^t W(s) ds}$. In this expression $W(t)$ is constant -- it's not a function of time, but the value of $W$ at the end of the integration period. We are therefore at liberty to move it under the integral. We're also at liberty, because of linearity, to move the expectation operator under the integral so that
\be
\ave{W(t) \int_0^t W(s) ds}=\int_0^t \ave{W(t) W(s)}ds
\ee
Of course, $W(t)=\int_0^t dW$, or in Langevin notation $W(t)=\int_0^t \eta(\tau) d\tau$. Similarly, $W(s)=\int_0^s \eta(\sigma) d\sigma$, and the noise functions $\eta(\tau)$ and $\eta(\sigma)$ are identical. In the ensemble of trajectories, we have $\ave{\eta(\sigma)\eta(\tau)}=\delta(\sigma-\tau)$, where $\delta$ denotes the Dirac delta function. Substituting all these pieces of information we find
\bea
\ave{W(t) \int_0^t W(s) ds}&=&\int_0^t ds   \int_0^t d\tau \int_0^s d\sigma \ave{\eta(\tau) \eta(\sigma)}\\
&=&\int_0^t ds   \int_0^t d\tau \int_0^s d\sigma \delta(\tau-\sigma)\\
&=&\int_0^t s ds  \\
&=& \frac{t^2}{2}.
\eea
Substituting in Eq.~\ref{eq:mean-square-2}, we finally arrive at
\be
\ave{\left(\frac{d\overline{W}_t}{dt}\right)^2}=\frac{1}{3t}.
\elabel{3t}
\ee
As suspected, the (root-)mean-square of the time derivative of the time average of a Brownian motion decays towards zero. Similarly the expectation value of the time variation decays to zero, 
\bea
\ave{v(t,\Delta t)}&=&\frac{1}{\Delta t}\int_{t-\Delta t}^t \frac{1}{3s}ds\\
&=& \frac{1}{3\Delta t}\ln\left(\frac{t}{t-\Delta t}\right),
\eea
approximately the same as Eq.~\ref{eq:3t} for small $\Delta t$.
Let's give this curious fact the appreciation it deserves: the variance (across the ensemble) of the time average $\overline{W}_t$ diverges with time as $\frac{t}{3}$, but the time-variation of same object vanishes as $\frac{1}{3t}$. 

This does not happen for ergodic observables. Think, for instance, of an observable described by an Ornstein-Uhlenbeck process. With weaker mean reversion, the ensemble of trajectories will be broader at each moment in time, and the time variation of a single trajectory will also increase. Such general features of stationary processes can shape our intuition, and that intuition can then fool us when we face non-ergodic, non-stationary processes like Brownian motion.

In our hypothetical scenario, the researcher will be misled if he assumes that a stationary value exists. Following our protocol, if the researcher were to run the experiment many times, he would wait approximately until $\ave{v(t; \Delta t)}$ has decayed to $\epsilon$, meaning  
\be
t^*\approx \frac{\Delta t}{1-\exp(-3\epsilon\Delta t)}
\ee
(roughly $1/(3\epsilon)$ for small $\epsilon \Delta t$) before concluding that sufficient averaging had taken place. No matter how small we choose $\epsilon$, the termination criterion will eventually be reached by any individual Brownian motion trajectory, not by chance but because the termination condition is systematically approached. Our straw-man researcher will incorrectly conclude that he has found the true value with a statistical error that is small enough to warrant publication. The published result, however, will not be reproducible.

\section{Numerical illustration}
\seclabel{Numerical}

We simulate our protocol applied to Brownian motion as follows. 

\begin{enumerate}
\item {\it Simulate a Brownian motion}\\
At time $t=0$ a trajectory starts at $W(0)=0$. We find $W(t+\delta t)$ by adding an instance of a normal random variable $\xi \sim \mathcal{N}(0,\delta t)$, 
\be
W(t+\delta t)=W(t)+\xi(t). 
\ee

\item {\it Compute the time average}\\
The time average of the trajectory is computed in each time step as
\be
\overline{W}_t=\frac{1}{t/\delta t+1}\sum_{n=0}^{t/\delta t} W(n\delta t). 
\ee

\item {\it Compute the time variation}\\
From time $t=\delta t$ onwards, the rate of change of the time average is computed as
\be
\frac{\delta \overline{W}_t}{\delta t}=\frac{\overline{W}_t-\overline{W}_{(t-\delta t)}}{\delta t}
\ee
From time $t=\Delta t+\delta t$ onwards, the time variation is computed as
\be
v(t,\Delta t) = \frac{1}{\Delta t/\delta t+1}\sum_{n=0}^{n=\Delta t/\delta t} \left(\frac{\delta \overline{W}_{t-\Delta t+n\delta t}}{\delta t}\right)^2
\ee

\item  {\it Termination}\\
When $v(t,\Delta t)<\epsilon$, the simulation is terminated.
\end{enumerate}

\begin{figure}[h!]
\begin{picture}(200,230)(0,0)
  \put(60,0){\includegraphics[width=.7\textwidth]{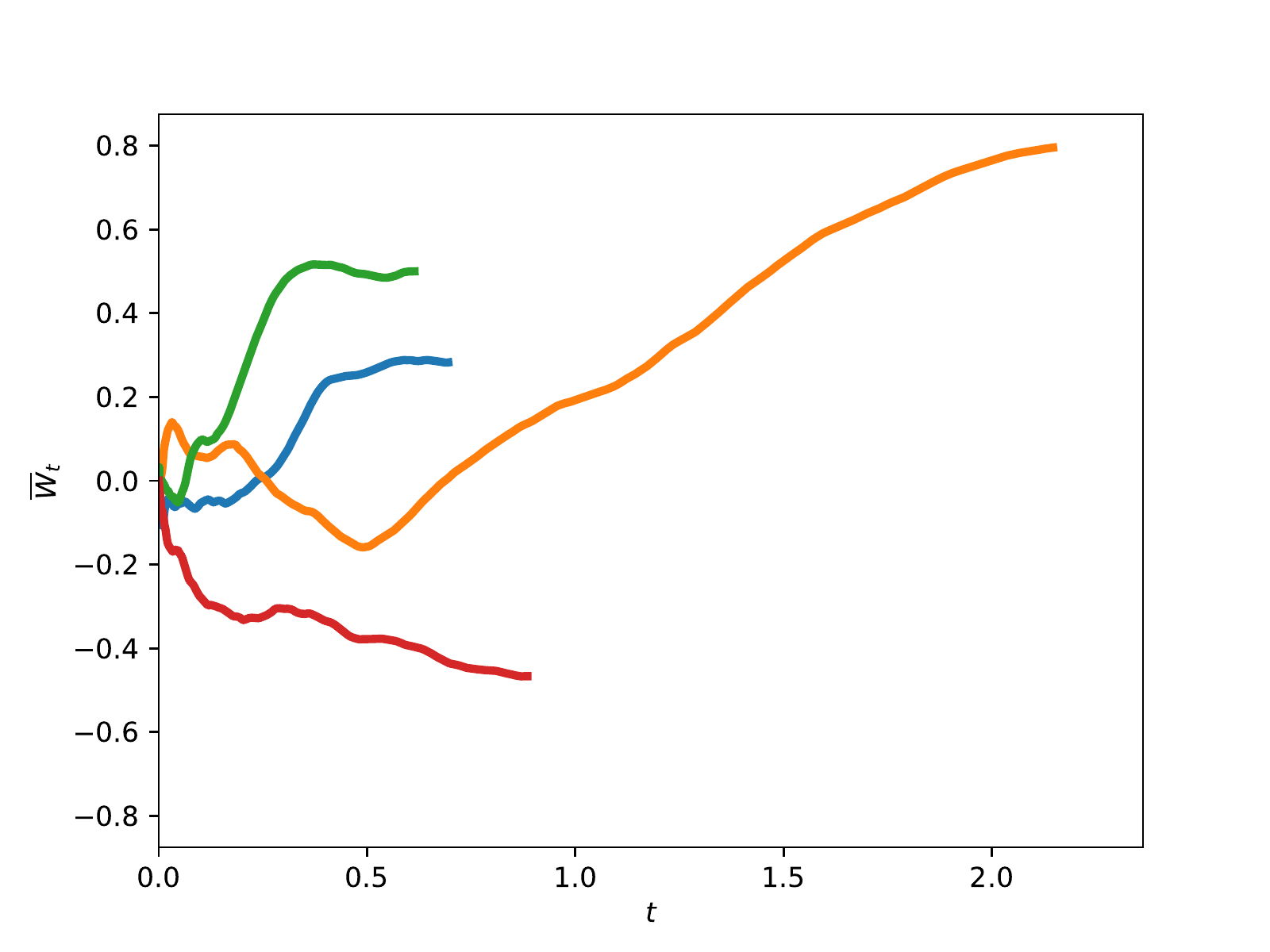}}
\end{picture}
\caption{Four repetitions of an experiment, showing realizations of $\overline{W}_t$ up to termination time. The stopping criterion is implemented using $\Delta t=0.3$ and $\epsilon=0.1$, so that $t^*\approx 3.5$. Eventually, the time average of each realization is stable according to the criterion. But the apparently stable values are different each time the experiment is repeated. The following features are observed. 1. Temporal behavior: each individual line becomes smoother over time. 2. Ensemble behavior: the width of the distribution (the range of values the different lines take) grows with time.}
\flabel{terminate}
\end{figure}

\section{Discussion}
\seclabel{Discussion}
Qualitatively, the message is clear: when I set out to measure the true value of some quantity, I typically assume that this value exists. The assumption of its existence alone can lead me to conduct an analysis that will look as if it does exist. However, often this will be wrong. The result is a measurement that seems stable in time but yields a different value each time it is repeated.

The essence of the problem is one of scaling. Time averaging involves two operations -- a division and a summation,
\be
\overline{a}_t=\underbrace{\frac{1}{t}}_{Division} \underbrace{\sum_i^t a_i}_{Summation}
\ee
An increasingly smooth time average suggests robustness of results, but it does not imply reproducibility. It is the scaling of the variance of the time average -- an ensemble property -- that tells us whether results are reproducible. This variance scaling is the product of the competing scalings of the two operations. The division always introduces a $T^{-2}$ scaling. The sum may scale in different ways. 
\begin{itemize}
\item
In the benign case -- think of $a_i$ as instances of a standard normal -- the variance of the sum scales as $T$. The variance of the time average $\overline{a}_t$ then scales as $T^{-2}\times T=T^{-1}$, which converges to zero, and measurements of $\overline{a}$ are reproducible.
\item
In the case we've studied here, the variance of the sum scales as $T^3$ (see eq.~\ref{eq:sum_dist}), and it dominates over the $T^{-2}$ scaling of the division. The variance of the time average $\overline{a}_t$ then scales as $T^{-2}\times T^{3}=T$. Measurements of $\overline{a}$, are apparently stable but not reproducible.
\item
Another pathological case that we have not studied here is that of heavy-tailed distributions, where the $a_i$ are instances of a random variable whose second moment does not exist -- think of the $a_i$ as instances of a Pareto distribution whose power-law tail implies a divergent second moment, \ie $P_a(x)\sim x^{-\tau}$ with $\tau<3$ for large $x$. 
The time average in this case will show jumps corresponding to extreme values sampled from the tail, and it will depend systematically on $T$.
\end{itemize}

Quantitatively, the story -- even just for the Brownian-motion example -- is not complete: according to our protocol termination occurs when $v(t; \Delta t)$ reaches $\epsilon$ {\it for the first time}, meaning that this is a first-passage problem. This explains the tendency in figure~\ref{fig:terminate} for trajectories to terminate before the time $t^*$. We have only computed at what time the expectation value of $v(t; \Delta t)$ will reach $\epsilon$. A further interesting and open question is how typical this expectation value is, \ie one might want to know the full distribution of $v(t; \Delta t)$. 

The key message is that our protocol, or any protocol like ours, can easily fool me into believing that a stable value has been measured. The error estimated from a single trajectory in an intuitively reasonable way will become arbitrarily small, but repeating the experiment will give a completely different value.

Our study highlights just how subtle the pitfalls of non-ergodic processes can be. A protocol to arrive at a good estimate for the value of a noisy but stationary observable (an ergodic process) was shown to be fooled by the simplest most common weakly non-ergodic process. Given that probability theory is much more commonly taught at universities than non-ergodic stochastic processes, it is likely that some irreproducible studies have inadvertently fallen into the trap we've described, namely by assuming ergodicity where it is not warranted.

\section*{References}
\bibliographystyle{iopart-num}
\bibliography{bibliography}
\end{document}